\documentstyle[12pt]{article}
\newcommand{\beq}{\begin{equation}}
\newcommand{\eeq}{\end{equation}}
\newcommand{\id}
 {i\kern.06em\hbox{\raise.25ex\hbox{$/$}\kern-.60em$\partial$}}
\newcommand{\as}{/\kern-.52em A}
\newcommand{\bs}{/\kern-.52em b}
\newcommand{\qs}{/\kern-.52em s}
\newcommand{\D}{{\cal{D}}}
\newcommand{\dv}{\!d^3\!x\,}

\newcommand{\dd}
{\kern.06em\hbox{\raise.25ex\hbox{$/$}\kern-.60em$\partial$}}

\newcommand{\tr}{\mathop{\rm tr}\nolimits}
\begin{document}
\title{A comment on bosonization in $d \geq 2$ dimensions}
\author{F.A. Schaposnik\thanks
	{Investigador CICBA, Argentina}\\
{\normalsize\it
Departamento de F\'\i sica, Universidad Nacional de La Plata,}\\
{\normalsize\it
C.C.~67, (1900) La Plata, Argentina.}
}
\date{}

\maketitle
\begin{abstract}
{We discuss recent results on bosonization in $d \geq 2$ space-time
dimensions by giving a very simple derivation for the
bosonic representation of the original free fermionic model both
in the abelian and non-abelian cases. We carefully analyse the
issue of symmetries in the resulting bosonic model as well as the
recipes for bosonization of fermion currents. }
\end{abstract}
\newpage
%===================================================================

It is only very recently that the powerful techniques of bosonization
of two-dimensional fermion systems \cite{bos} have been extended
with some success to $ d >2 $ space-time dimensional models
\cite{bur}-\cite{fs1} (see refs.\cite{ref} for previous efforts in
this direction).
In particular, the well known 2-dimensional abelian bosonization recipe
for the fermionic current in terms of a bosonic field $\phi$,
\beq
\bar \psi \gamma^{\mu} \psi  \to {({1}/{\sqrt{\pi}})}
\epsilon^{\mu \nu} \partial_\nu \phi
\label{1}
\eeq
has been shown to be generalizable at least in the
case of $ d = 3$ abelian
fermionic models to the formula \cite{fs}
\beq
\bar\psi \gamma^{\mu} \psi  \to \pm i{\sqrt \frac{1}{4\pi}}
\epsilon^{\mu \nu\alpha}
\partial_{\nu}A_{\alpha}
\label{aq1}
\eeq
where $A_{\alpha}$ is a gauge field with Chern-Simons
dynamics and the
recipe is only valid to order $1/m$
($m$ being the fermion mass).
This result, already implicit in ref.\cite{bur},
has been extended to the
case of free non-Abelian fermions in $d=3$ dimensions (again to order $1/m$)
\cite{fs1},
\beq
\bar\psi \gamma^{\mu} t^a\psi  \to \pm i{\sqrt \frac{1}{4\pi}}
\epsilon^{\mu \nu\alpha}
F_{\nu\alpha}^a
\label{aq}
\eeq
where $F_{\nu\alpha}$ is the field strength of a non-abelian
gauge field with Chern-Simons dynamics. As explained in
\cite{fs1}, formula (\ref{aq}) is, in some sense,
 the 3-dimensional analogue of
the 2-dimensional non-abelian fermion-boson mapping
\beq
j_+ \to -\frac{i}{4\pi} h^{-1}\partial_+ h
\label{w1}
\eeq
\beq
j_- \to -\frac{i}{4\pi} h \partial_- h^{-1} .
\label{w2}
\eeq
Although the approaches of refs.\cite{bur}-\cite{fs1} have many points
in common and lead to equivalent results, the road they follow to
attain bosonization is somehow diverse. For example,
the results in
refs.\cite{fs}-\cite{fs1} are based on the construction of the
master or interpolating bosonic Lagrangian introduced in
refs.\cite{TPv}-\cite{LPvT} for studying self-dual
systems and seems to be the most appropriate  for
generalizations to non-abelian systems (at least in $d=3$). The approach of
ref.\cite{bur} exploits
the connection between bosonization and duality transformations discovered
in \cite{bur1} and leads in a very elegant way to the abelian boson-fermion
mapping; it seems to be the most adequate for studying $ d > 3$ models.

It is the purpose of this note to comment on the connection
between
these different proposals, showing their common origin
through the
analysis of an alternative way of deriving the fermion-boson
mapping.
As we shall see, this alternative approach is related
with a priori
disconnected ideas, as those in Faddeev-Shatashvili
proposal of introducing new gauge degrees of freedom for the
consistent quantization of anomalous gauge theories \cite{FadS}-\cite{HT}
or in the "smooth" bosonization of certain 2 dimensional models
\cite{dns}-\cite{tssg}.

We start from the (Euclidean) Lagrangian for free massive Dirac fermions
in $ d$ dimensions
\beq
{\cal L}_{F}= \bar\psi (\id + m) \psi
\label{L}
\eeq
where fermions are in the fundamental representation of
some group $G$.
In most cases, we shall consider  $G = U(N)$ so that the
conserved fermion current associated
with Lagrangian (\ref{L}) reads
\beq
j^{a\mu} = \bar\psi^i t^a_{ij} \gamma^{\mu} \psi^j
\label{jn2}
\eeq
with $t^a$ the $U(N)$ generators.

This current stems from the {\underline{global}} $U(N)$
invariance of
(\ref{L}) under the transformation
\[
\psi \to g \psi
\]
\beq
\bar \psi \to \bar \psi g^{-1}
\label{cha}
\eeq
with $g$ an element of $U(N)$. At the quantum level, current conservation
can be derived using Noether method by starting from the
partition function

\beq
Z_{F} = \int \D\bar\psi \D\psi \exp[
-\int  \bar\psi (\id + m) \psi d^dx],
\label{2}
\eeq
promoting $g$ to a local transformation,
\[
\psi \to g(x) \psi
\]
\beq
\bar \psi \to \bar \psi g^{-1}(x)
\label{chas}
\eeq
and then taking (\ref{chas}) as a change of the fermionic
variables
in $Z_F$.
After considering $g(x)$ infinitesimally close to the identity
 one straightforwardly obtains
\beq
< \partial_\mu j_\mu ^a > = 0 .
\label{Noether}
\eeq
Now, the local transformation (\ref{chas}) plays a central
role in our
route to bosonization. Indeed, our procedure starts by taking
(\ref{chas}) as  a (finite) change of variables in (\ref{2}),
\[
\psi =g(x) \psi'
\]
\beq
\bar \psi = \bar \psi' g^{-1}(x) .
\label{chas2}
\eeq
One can always define for Dirac fermions a path integral
measure invariant under
transformation (\ref{chas2}). Then, after the change of variables
the partition function becomes
\beq
Z_{F} = \int \D\bar\psi \D\psi \exp[
-\int  \bar\psi (\id + m +i g^{-1}\dd g) \psi d^dx].
\label{n2}
\eeq
(we have omitted primes in the new fermionic variables).
Being $Z_{F}$ $g$-in\-de\-pen\-dent, we can integrate both sides
in eq.(\ref{n2})
over $g$ using a Haar measure $D g$,
this amounting to a trivial change in the normalization of the path-integral.
In performing this integration we include an
arbitrary weight ${\hat F}[g]$ which will play an important
role in what
follows \footnote{The suggestion of introducing this functional
and its relevance concerning symmetries was done by N.Brali\'c \cite{Bral}.}.
We then have
\beq
Z_{F} = {\cal N} \int \D \bar\psi \D \psi \D g {\hat F}[g]\exp[
-\int  \bar\psi (\id + m +i g^{-1}\dd g) \psi d^dx].
\label{n222}
\eeq
It is evident that $i g^{-1}\dd g$  in (\ref{n222}) can be
thought as a flat connection and can be then replaced by a
"true" gauge field connection provided a constraint is
introduced
to assure its flatness. Indeed,  using the identity

\beq
\int \D g {\cal{H}}[i g^{-1}\dd g] =
\int \D b_{\mu} {\cal{H}}[b]
\delta[\epsilon_{{\mu_{1}} {\mu_{2}}...{\mu_{d}}}
 f_{{\mu_{1}}{\mu_{2}}}]
\label{id}
\eeq
with
\beq
f_{\mu \alpha} = \partial_\mu b_\alpha - \partial_\alpha b_\mu
+ i[b_\mu,b_\alpha]
\label{nf}
\eeq
we can rewrite (\ref{n222}) in the form
\beq
Z_{F} = \int \D \bar\psi \D\psi \D b_\mu F[b]
\delta[\epsilon_{{\mu_{1}} {\mu_{2}}...{\mu_{d}}}
 f_{{\mu_{1}}{\mu_{2}}}]
\exp[-\int  \bar\psi (\id + m + \bs) \psi d^dx].
\label{ntrade}
\eeq
The next step in our derivation is to integrate out
fermions so that
the partition function becomes:

\beq
Z_{F} = \int Db_\mu F[b] det(\id + m + \bs)
\delta[\epsilon_{{\mu_{1}} {\mu_{2}}...{\mu_{d}}}f_{{\mu_{1}}{\mu_{2}}}] .
\label{in}
\eeq
The central point in our bosonization route is now at sight:
whether one would arrive
to an exact bosonization formula or to an approximate recipe will depend
on the possibility of computing the fermion determinant in a
closed form.
If this were possible, then the fermionic degrees of freedom,
which
disappeared
from the partition function would  be replaced by new bosonic degrees of
freedom,
in an exactly equivalent bosonic model. Otherwise, bosonization
will
be approximate.

As it is well-known, in the $d=2$ massless case,
the fermion determinant can be computed exactly
both in the abelian and non-abelian cases and then  exact
bosonization rules can be trivially obtained from (\ref{in}). In $d=3$
the fermion determinant can be computed as a $1/m$ expansion this
leading to a fermion-boson mapping valid at large-distances. This
and other approximation methods would lead to  (approximate)
bosonization rules in higher dimensions.

In order to proceed, one has to exponentiate the delta function in
(\ref{in}) by means of a Lagrange multiplier field. As explained above,
it is this
bosonic field, a scalar in $d=2$, a vector in $d=3$, an antisymmetric
(Kalb-Rammond) field in $d > 3$ dimensions, which will play the
central role in the bosonized theory, once the auxiliary
$b$-field
is integrated out. This will become clear in the following simple
examples:
%\newpage

\noindent {\underline{The $d=2, U(1)$, massless case}}
\vspace{0.3 cm}

In this case we can use the well-known result:
\beq
\log det(\id + \bs) = - \frac{1}{2\pi}\int d^2x
b_{\mu}(\delta_{\mu \nu} - \partial_{\mu} \Box^{-1} \partial_{\nu})b_{\nu} .
\label{det2}
\eeq
This, together with the representation
\beq
\delta[\epsilon_{\mu \nu}f_{\mu \nu}] = \int D\phi \exp(-\frac{1}{\sqrt{\pi}}
\int d^2x \phi \epsilon_{\mu \nu}f_{\mu \nu})
\label{lag2}
\eeq
leads, after a trivial gaussian integration over $b_\mu$, to the result:
\beq
Z_F = \int D\phi \exp(-\frac{1}{2}\int d^2x
\partial_{\mu}\phi \partial_{\mu}\phi)
\label{Z}
\eeq
relating the free fermion partition function $Z_F$ with the partition
function of the corresponding free boson field.
Notice that in arriving to eq.(\ref{Z}) we have fixed
 the arbitrary function $F[b] = 1$. As we shall see
below, the choice of $F[b]$ is related to the question of gauge invariance
of the resulting bosonic theory and  it is in the
non-abelian case where  a non-trivial choice of $F[b]$
becomes important.

Concerning bosonization rules for fermion currents, let us
note that the addition of a fermion source
$s_{\mu}$ in $Z_F$ amounts to the inclussion of
this source in the fermion determinant
\beq
Z_{F}[s]= \int Db_\mu  det(\id + \bs + \qs)
\delta[\epsilon_{\mu\nu} f_{\mu \nu}] .
\label{in2}
\eeq
Now, a trivial shift $ b + s \to b$ in the integration
variable $b$ puts
the source dependence into the constraint
\beq
Z_{F}[s]= \int Db_\mu  det(\id + \bs)
\delta[\epsilon_{\mu\nu} (f_{\mu \nu} - 2\partial_\mu s_\nu)]
\label{in22}
\eeq
so that, instead of (\ref{Z}) one ends with
\beq
Z_F[s] = \int D\phi \exp (-\frac{1}{2}\int d^2x
(\partial_{\mu}\phi \partial_{\mu}\phi + \frac{2}{\sqrt
\pi}s_\mu
\epsilon_{\mu \nu} \partial_\nu \phi)) .
\label{Z2}
\eeq
By simple differentiation with respect to the source
one infers from this expression the bosonization recipe for
$j_\mu$ given in eq.(\ref{1}).
\vspace{ 0.5 cm}

\noindent {\underline{The $d=3, U(1)$, massive case}}
\vspace{0.3 cm}

The fermion determinant in $d=3$ dimensional space-time
cannot be computed
in a closed form. One can however consider an approximation
approach which
in the present case can be envisaged as an expansion in
inverse powers
of the fermion mass, $1/m$. Indeed, following refs.\cite{DesJ}-\cite{GMSS} one
gets an expression containing parity violating contributions as
well as parity conserving terms which can be computed order
by order
in $1/m$
\beq
 \ln \det (\id+ m +\bs) = \pm \frac{i}{16\pi }
\int\epsilon_{\mu\nu\alpha} f^{\mu\nu} b^{\alpha} \dv + I_{PC}[b_{\mu}]+
O(\partial^2/m^2)
,
\label{9f}
\eeq
\beq
 I_{PC}[b_{\mu}] = - \frac{1}{24\pi  m} \int \dv  f^{\mu\nu} f_{\mu\nu}
+ \ldots
\eeq
Using this result to the leading order in $1/m$,
the corresponding partition function  $Z_F$ can be written in the form
\beq
Z_{F} \simeq \int Db_\mu DA_\mu
\exp[-\int \left(\mp
\frac{i}{8\pi}\epsilon^{\mu\alpha\nu}
b_{\mu}\partial_{\alpha}b_{\nu}  +
A_\mu\epsilon_{\mu \nu \alpha}f_{\nu \alpha} \right )\dv]
\label{extra}
\eeq
where $\simeq$ indicates that the identity is valid to the
lowest
order in $1/m$.
Again, the integration over $b_\mu$ is quadratic so that it
can be
trivially performed so that one finally obtains
\beq
Z_{F} \simeq \int  DA_\mu \exp[\pm \frac{i}{2}\int \dv \epsilon^{\mu\alpha\nu}
A_{\mu}
\partial_{\alpha}
A_{\nu}]
\label{cosa}
\eeq
or
\beq
Z_{F} \simeq Z_{CS}
\label{equ}
\eeq
where $Z_{CS}$ denotes the partition function for a
pure Abelian Chern-Simons theory.
Equation (\ref{equ}) establishes the connection between
a theory of free
fermions and the pure Chern-Simons theory in $2+1$ dimensions,
to the lowest order
in inverse powers of the fermion mass. This last restrictions implies
that our results are valid only for long distances in contrast
with
$1+1$ bosonization which is in a sense a short-distance result.
This
peculiarity makes the comparison of the commutator algebra, which tests
short distances, somehow hazardeous.

As in the previous example, we could have added an external
source $s_{\mu}$
for the fermion current
in the original fermionic Lagrangian. After the trivial shift
$b_\mu  + s_\mu \to b_\mu$
in (\ref{ntrade}) we end with an identity of
the type (\ref{cosa})-(\ref{equ}) but now in the presence of sources:
\begin{eqnarray}
Z_{F}[s] &=& \int \D\bar\psi \D\psi \exp[
-\int \left( \bar\psi (\id + m) \psi + s_\mu j_\mu \right)\dv] \nonumber \\
& \simeq & \int  DA_\mu \exp[\pm \frac{ i}{2}\int  (\epsilon^{\mu\alpha\nu}
A_{\mu}
\partial_{\alpha}
A_{\nu} + \sqrt{2/\pi}
s_\mu\epsilon^{\mu\alpha\nu}\partial_{\alpha}
A_{\nu}) ~\dv] .
\label{2d}
\end{eqnarray}
{}From this, we confirm the bosonization rule for the fermion
current given by eq.(\ref{aq1}).
\vspace{0.3 cm}

\noindent \underline{Symmetries}

Let us discuss at this point
the issue of symmetries in our bosonization approach. We have started
from a free fermionic Lagrangian invariant under global $U(N)$
rotations. Through transformations (\ref{chas2}), which
are the local counterpart of those global rotations, we have
forced the appearence of  gauge degrees of freedom in the
effective Lagrangian.
Although trivial at the start (they enter through a flat
connection) these
degrees of freedom become non-trivial once flatness is
implemented
via a Lagrange multiplier. So, the bosonic equivalent
of the original fermionic Lagrangian is a Lagrangian for a
Lagrange
multiplier whose character (scalar, vector, in general a rank
$d-1$
completely antisymmetric field which in the dualization
approach to bosonization is taken as a Kalb-Ramond gauge
potential
\cite{bur}) depends on the
space-time dimensionality. How does the imposed local symmetry reflect
in the resulting bosonic model?
The bosonic field $\phi$ enters in our approach through the
delta function ensuring flatness,
\beq
\delta[\epsilon_{{\mu_{1}} {\mu_{2}}...{\mu_{d}}}f_{{\mu_{1}}{\mu_{2}}}]
 = \int D\phi_{} \exp(-\frac{1}{\sqrt{\pi}}
tr\int d^dx \epsilon_{{\mu_{1}} {\mu_{2}}...{\mu_{d}}}
\phi_{{\mu_{3}}...{\mu_{d}}}f_{{\mu_{1}}{\mu_{2}}} ) .
\label{lag22a}
\eeq
Now, since the l.h.s. in this identity is gauge invariant, one
should define the r.h.s. accordingly.
In the two abelian examples discussed above, the curvature
$ f_{{\mu_{1}}{\mu_{2}}}$ is gauge-invariant and then  (\ref{lag22a})
does not force the Lagrange multiplier to have a definite transformation
law.
In particular, in the $2$-dimensional
case $\phi$ is just a scalar which does not partake of local
gauge transformations. In the $3$-dimensional case, it is
reasonable to interpret $\phi_\mu$ as a gauge connection
$ \phi_\mu \equiv A_\mu$ since the resulting effective bosonic action
is (to order 1/m) a Chern-Simons action for $ A_\mu$ which
naturally possesses a gauge-invariance.

The issue is more subtle,
 in the non-abelian case, where $f_{\mu \nu}$ changes
covariantly.
One could in principle demand
$ \phi_{{\mu_{3}}...{\mu_{d}}}$  to change
covariantly in order to have a gauge invariant r.h.s. . In the
$d=2$ case, this problem can be overcome just by fixing
the gauge degrees of freedom associated with $b_\mu$, as
discussed in \cite{dOQ},\cite{tssg}. In particular, in this
last reference, in an approach similar to the one here
presented,
non-abelian two-dimensional bosonization is discussed in the
spirit of smooth bosonization \cite{dns}.

Concerning higher dimensions, already for
$d=3$ the problem is more complicated. It is in handling
this problem that  the
arbitrary functional $\hat F[g] = F[b]$ appearing
in eq.(\ref{in}) finds its relevance: since the
$d=3$ non-abelian bosonization recipe
summarized by eq.(\ref{aq})  leads to a bosonic
model with  non-abelian Chern-Simons dynamics,
one should expect the bosonic partition function to possess
gauge invariance.
With this in mind,  one can fix $F[b]$ so
that the bosonic field $\phi_\mu$ can be  again taken as a
gauge connection,
this time taking values in the Lie algebra of $U(N)$ and the resulting theory
then
exhibits gauge-invariance.
This can be achieved by chosing
$F[b]$ in the form

\beq
F[b] = \exp[ \frac{i}{12 \pi} \tr \int \dv
\epsilon_{\mu \nu \alpha} b_\mu
b_\nu b_\alpha ]
\label{elec}
\eeq
Indeed,  under gauge transformations
\beq
b_\mu \to b_\mu^h = h^{-1} b_\mu h + i h^{-1} \partial_\mu h
\label{gauge}
\eeq
$F[b]$ then changes as
\beq
F[b] \to F[b^h] = F[b] \times \exp (2 \pi i \omega[h]) \times
\exp [\frac{i}{16\pi} \tr \int \dv \epsilon_{\mu \nu \alpha}
f_{\mu \nu} h
\partial_\alpha h^{-1}]
\label{cis}
\eeq
with $\omega[h]$ the winding number of $h$,
\beq
\omega[h] = \frac{1}{24 \pi^2} \tr \int \dv \epsilon_{\mu \nu \alpha}
h^{-1}\partial_\mu h h^{-1}\partial_\nu h
h^{-1}\partial_\alpha h .
\label{o}
\eeq
For compact non-abelian gauge groups (thus with $\Pi_3 = Z$) $\omega[h]$
takes integer values so the second factor in the r.h.s. of (\ref{cis}) is
irrelevant.
Concerning the last one, it precisely cancels out that arising
from the delta function representation when the Lagrange
multiplier
$ \phi_\mu \equiv A_\mu$ changes as a connection
\beq
A_\mu \to A_\mu^h = h^{-1} A_\mu h + i h^{-1} \partial_\mu h .
\label{gauge1}
\eeq
That is, one has
\beq
F[b^h] \times \exp(-\frac{1}{\sqrt{\pi}} tr \int d^3x
\epsilon_{\mu \nu \alpha} f_{\mu \nu}^h A_{\alpha}^h )
=
F[b] \times \exp(-\frac{1}{\sqrt{\pi}} tr \int d^3x
\epsilon_{\mu \nu \alpha} f_{\mu \nu} A_{\alpha})
\label{inv}
\eeq

In order to summarize what we have learned in the
three dimensional case concerning symmetry, let us
write the partition function (\ref{2}) in terms of the bosonic
field $ A_\mu$, once
 bosonization
is achieved, in the form

\beq
Z_F = \int D A_\mu \times \exp(- S_B[A_\mu])
\label{lin}
\eeq
with
\beq
\exp(-S_B[A_\mu])
=
\int D b_\mu F[b] det(\id + m + \bs)  \exp(-\frac{1}{\sqrt{\pi}}
tr\int d^3x
\epsilon_{\mu \nu \alpha} f_{\mu \nu} A_\alpha )
\label{lind}
\eeq
Now, using eq.(\ref{inv}) as well as the invariance of the
fermion determinant and the path-integral measure under
the change $b_\mu \to b_\mu ^h$, one trivially sees that
\beq
\exp(-S_B[A_\mu ^h]) =
\exp(-S_B[A_\mu]) .
\label{lindo}
\eeq
That is, the bosonic theory can be endowed, in 3 dimensions,
with a local
gauge symmetry and this is an exact result in the sense it does
not arise
from some approximation for the fermion determinant. Of course,
the role
of the functional $F[b]$ is crucial for this property.
Presumably,  there is always a choice for $F[b]$ such that,
as
it hapens in $d = 3$,
the $d$-dimensional bosonic effective theory  can be
defined as a theory in which the bosonic  field
$\phi_{{\mu_{3}}...{\mu_{d}}}$ is a Kalb-Ramond
\underline{gauge} field.

Equation (\ref{lindo}) can be seen as connecting the approach presented
here with Faddeev-Shatashvili proposal for quantizing
anomalous theories
\cite{FadS}. Indeed, as explained in \cite{BSV}, this
proposal can
be implemented by starting from a partition function where (classically
redundant) gauge degrees of freedom are taken into account at
the quantum level. It is in this way
that gauge invariance of the effective action for chiral
fermions is
achieved and an identity analogous to (\ref{lindo}) can
be proved. From
this, consistency of anomalous gauge theories can
be proved,
at least in the two dimensional case.

The analysis above shows how our approach, inspired in the bosonization
proposal
of ref.\cite{bur}, sheds light on the issue of gauge-invariance of the
resulting bosonic theory. This becomes a central point
when deriving
$3$-dimensional non-abelian bosonization. Indeed, up to
date, only the
approach of refs.\cite{fs}-\cite{fs1}, based in the
construction of an
auxiliary master Lagrangian has led to a concrete recipe
for
bosonization of the fermionic
current (eq.(\ref{aq})). This recipe shows that, for
large distances, the bosonic theory is endowed with a
local gauge
invariance and the precedent discussions clarifies why
this is so. The
derivation of the bosonization recipe (\ref{aq})
within the approach
presented here is difficulted by the fact that non-linear
terms
prevent the trivial integration of the auxiliary
field $b_\mu$. This
problem can be handled in the same way it was done in the
interpolating Lagrangian approach \cite{fs1} but we hope that a
three dimensional analogous of Polyakov-Wiegman identity
\cite{Pol}
may simplify
the derivation following the lines of the present work.

The construction of the interpolating Lagrangian in \cite{fs}-\cite{fs1}
heavily depends on the fact that, for $d=3$,
a connection between selfdual and Maxwell-Chern-Simons
models can be
established \cite{DJ}. For higher di\-mensions, we think
that  the method
of ref.\cite{bur}, pre\-sented he\-re through an alternative
route,
is better
adapted. One should note in particular
 the simplicity with which one can include sources for the
fermion
current since, after a shift in the auxiliary field $b$, the
source $s_\mu$ always ends coupled to the bosonic field in the
form

\beq
S_{source}[s] = tr\int d^dx
\epsilon_{{\mu_{1}} {\mu_{2}}...{\mu_{d}}}s_{\mu_{1}}
\partial_{\mu_{2}}\phi_{{\mu_{3}}...{\mu_{d}}} +O(s^2)
\label{lag21}
\eeq
this suggesting a bosonization rule of the form

\beq
\bar \psi \gamma_{\mu_{1}} t^a \psi \to \epsilon_{{\mu_{1}}
{\mu_{2}}...{\mu_{d}}}\partial_{\mu_{2}}
\phi^a_{{\mu_{3}}...{\mu_{d}}} .
\label{lance}
\eeq
Of course, the construction of the actual bosonic model in which this
fermion-boson mapping is realized will depend on the
posibility of evaluating the fermionic determinant.
For $d \geq 3$ one can only envisage approximations
(like for example the $1/m$
expansion employed in the $d=3$ case) but the basic steps
leading to
bosonization, summarized by  eqs.(\ref{chas2})-(\ref{in}),
remain the same
no matter the number of space-time dimensions.
\vspace{1 cm}

\underline{Acknowledgements}: The author wishes to thank
Ninoslav Brali\'c, Daniel Cabra and Eduardo Fradkin
for very helpful comments and suggestions which, after many discussions,
motivated this work.

%%%%%%%%%%%%%%%%%%%%%%%%%%%%%%%%%%%%%%%%%%%%%%%%%%%%%%%%%%%%%%%%%%%%%%%%%%
\end{document}